\documentclass[12pt]{book}

\usepackage[dvips]{graphicx,color}
\usepackage{makeidx,tsukuba}

\makeauthorindex
\makeindex

\begin{document}

\BookTitle{\itshape The 28th International Cosmic Ray Conference}
\CopyRight{\copyright 2003 by Universal Academy Press, Inc.}
\pagenumbering{arabic}

\chapter{A Proportional Wire Chamber Array: {\small GRAND}'s Status}
\author{John Poirier, Christopher D'Andrea, Mari\`e L\`opez del Puerto,\\
Eric Strahler, and Jon Vermedahl\\
{\it
Center for Astrophysics at Notre Dame, University of Notre Dame,\\
Notre Dame, IN 46556, USA}}

\section*{Abstract}

Project {\small GRAND}  is a 100m $\times$ 100m air shower array of
position sensitive proportional wire chambers
(PWCs) located at $41.7^{\circ}$ N and  $86.2^{\circ}$ W at an elevation of 220~m
above
sea level.  Its convenient location
adjacent to the campus of the University of Notre Dame
makes it a good training ground for students.
There are 64 stations each with eight 1.29 m$^2$ PWCs.  The
geometry of the stations allows for the angles of charged secondaries to be
determined to within 0.26$^\circ$ in each of two orthogonal planes; muons are
differentiated from electrons and hadrons by means of a steel plate.
Two triggers are run simultaneously:
a multiple hut coincidence trigger, rich in extensive air showers, and
a single track trigger, rich in secondary muon tracks.
The former trigger is sensitive to primary energies $\ge$100 TeV,
the latter to energies $\ge$10 GeV.



\section{Introduction}
Each station of {\small GRAND} has eight planes, four measure track positions
in the x (or east) direction, four in the y (or north) direction.
The four pairs of x,y planes are placed vertically (z) above each other with a 50 mm steel 
plate above the bottom pair of planes.  This plate allows muons (which pass
undeflected through the plate) to be differentiated from hadrons and
electrons (which are stopped, shower, or deflected by the steel plate).
A series of hits which
fit a straight line in the xz-plane without nearby hits in the bottom x-plane
is called a muon (in the x-view); any other pattern of hits is called an
electron or hadron in that view.
The yz-plane is analyzed independently of the xz-plane
with no attempt to correlate a track between the xz- and yz-planes.
The array is operated continuously with 96\% running efficiency.
It utilizes two simultaneous triggers:
1) a shower trigger, rich in extensive air showers (rate of $\sim$1 Hz), and
2) a single track trigger, rich in muon tracks (rate of 2400 Hz).
The absolute time for the triggers is recorded to one millisecond precision
using a clock synchronized to  WWVB.
A shift register memory on each PWC plane stores the
status of every wire from the last station trigger until overwritten by another
track(s) or the data are read out.

\section{Shower Trigger for Extensive Air Showers}

The shower trigger is: a hardware coincidence (three stations in time coincidence)
and an on-line software requirement (four stations tagged as coincident).
The status of each of the 40,960 wires is read into the master computer and
then written out to a separate 8 mm Exabyte tape drive (with compression) with
no on-line pre-analysis.  Since the 1~Hz data rate is mainly composed of
sparse data hits, a compressed magnetic tape can handle several
weeks of continuous data.  A histogram of the number of charged tracks which reach
detection level from an extensive air shower (EAS) shows the knee of the cosmic
ray spectrum [8].
The primary composition has been studied in a region around the knee by
tracing the muons back (upward) to find their origin.
Since the muons come primarily from the hadronic (upper) part of the EAS, the height of 
the muons is correlated with the first interaction of
the EAS which is a measure of the cross section (and therefore the identity) of the 
primary.
There is an indication that the cosmic ray primaries become heavier at and
above the knee region [8].
The CORSIKA code is being used to calculate {\small GRAND}'s response to cosmic rays
above the 100 TeV range of energies.

\section{Single Track Trigger for Muons}

The single track trigger is an 800 Hz oscillator
providing an artificial trigger which collects all the stored track data in
every station.
The 40,960 wires of the entire array data are read into the data
acquisition system in seven microseconds.
Eight on-line computers search through this data and select those stations
which have one and only one hit in each of the eight planes of that station
(two adjacent hits are also allowed, as there is a 28\% chance that a single
track will cause two adjacent hits).
Upon finding such a station, the numbers of the hit wires are recorded as a candidate
muon in the buffer memory of that computer.
After 900 such candidates are stored, the buffer is written to an
8~mm magnetic tape drive.
In the off-line analysis of this data, 96\% fit a straight line; from MC
calculations and 96\% of the straight lines are muon tracks.  Further details on the
construction and operation of {\small GRAND} as well as listings of prior publications can 
be found in [9,10].

The FLUKA Monte Carlo code is utilized to predict {\small GRAND}'s response to
primary hadrons and gamma rays.
The results for gamma rays are contained in [2,11]
and those for protons in
[6].
Assuming a primary spectral index of 2.4, the peak sensitivity is
for a primary energy of $\sim$10 GeV, whether it be a gamma ray or hadronic
primary.
Gamma ray primaries have a detectable signal of secondary muons from
gamma-hadro production in the atmosphere.  FLUKA calculations show that a TeV
gamma ray striking the atmosphere at normal incidence produces 0.23 muons which
reach {\small GRAND} [14].  Thus,
paradoxically,
single muons in {\small GRAND} can be
used as a $signature$ for gamma ray primaries.

\begin{figure}[t]
  \begin{center}
    \includegraphics[width=5.25in]{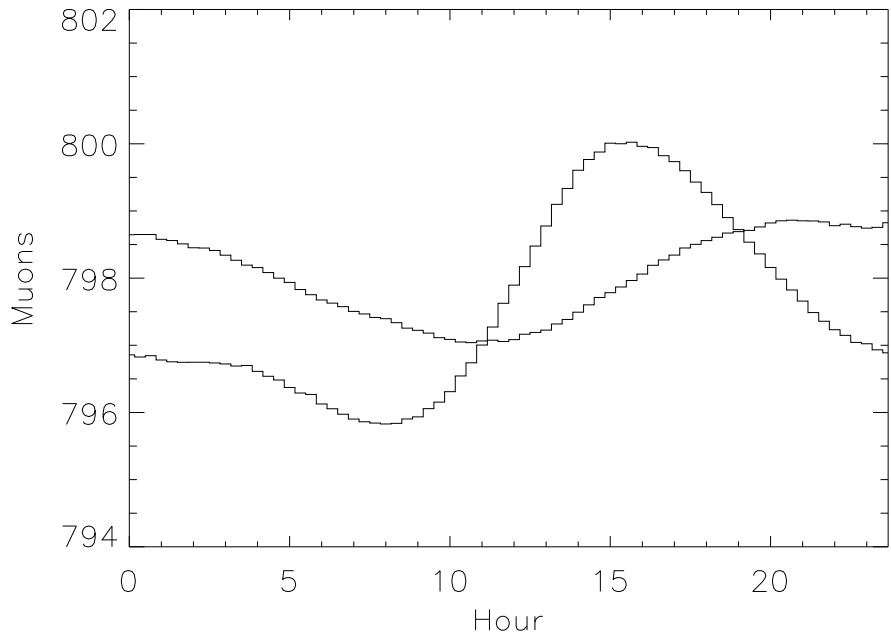}
   \end{center}
        \vspace{-0.3in}
  \caption{This figure shows the variation by right ascension (distribution with the 
smaller variation) and by hour of solar day (distribution with the larger variation).
The figure shows the top 1\% of the muon counting rate data.
The ordinate is millions of muons per 1/3 hour.}
	\vspace{-0.1in}
 \end{figure}

Some of the results from the single track trigger are discussed below.
The feature of using single muons as a signature for gamma ray primaries has been
used to
look for coincidences
between GRBs as observed by BATSE and gamma ray showers in the 10 GeV $\le$
E$_\gamma$ $\le$ 1 TeV energy region.
A search among eight candidates achieved a 2.7$\sigma$ deviation from background
for the one event which was, a priori, judged to be the most likely to be observed
[11,12].
{\small GRAND}'s ability to correlate short bursts of muons with an
identifiable source of primary cosmic radiation has been shown in a detection
which was coincident with a solar flare on 15 April 2001.
The statistical significance of this observation was at the
level of 6$\sigma$ for a ground level event of 0.6 hours duration [1,6].

The single muon data are used to generate files containing information on
the number of muons originating from each 1$^\circ\times1^\circ$ of right
ascension and declination during a complete sidereal day.
We have previously analyzed these data for possible asymmetries in the muon angles
[3,4,5,7,13]  We update this analysis below. In order to eliminate possible spurious 
variations in counting rate caused by experimental problems (such as a station being 
offline for repair during part of that day),
a smoothness test is imposed on the data files of each day.
First, the number of muons detected is summed over all
declinations for each degree of right ascension and the average and standard
deviation ($\sigma$) of the values calculated.
If the ratio of $\sigma$/average is greater
than 0.03, then that day's data file is not used.
From January 1997 through February 2003, 1665 data files were collected.
After the 3.0\% smoothness cut, there remained
1295 data files containing information on 101 billion muons.

Although Project {\small GRAND} has an average muon angular resolution of
$0.26^\circ$, secondary muons themselves have a birth-angle relative to
the primary and are further scattered and deflected as they traverse the
atmosphere and geomagnetic field.  This degrades the resolution for the primary
to about $\pm3^\circ$ - $5^\circ$ (depending on energy).  The muon information is
summed
over all declinations and $5^\circ$ bins of right ascension.
Figure 1 shows the number of counts per five degrees of right ascension.
The distribution in right ascension is extremely flat;
the maximum variation (from min to max) is 0.23\% of the total
counting rate.
The curve with the maximum variation (0.55\%) is that for hour-of-solar-day
and is more than twice as large as the sidereal variation.
The data are extremely isotropic in both right ascension and hour-of-solar-day,
though with large statistics it is possible to see small variations from isotropy.
The largest variation has to do with solar effects, probably the magnetic field
of the sun.
There is a residual asymmetry due to sidereal effects, possibly due to the
galactic magnetic field.

The authors wish to acknowledge contributions of
D. Baker, J. Gress, G. Terry, and G. VanLaecke.
Project {\small GRAND}'s research is presently being funded through a grant
from the
University of Notre Dame and private grants.
\section{References}
%
%

\re
\ 1. D'Andrea C., Poirier J.  \ 2003, Proceedings of the 28th ICRC

\re
\ 2. Fass\`o A. and Poirier J., \ 2000, Physical Review D 63, 036002

\re
\ 3. Lin T.F., \ et al., \ 1999, Proceedings of the 26th ICRC 2, 100

\re
\ 4. Poirier J., D'Andrea C., \ 2001, Proceedings of the 27th ICRC, 3923

\re
\ 5. Poirier J., D'Andrea C. \ 2001, Proceedings of the 27th ICRC, 3934

\re
\ 6. Poirier J., D'Andrea C. \ 2002, Journal of Geophysical Research 107, 1376

\re
\ 7. Poirier J., D'Andrea C., Dunford M., \ 2001, Proceedings of the~27th ICRC,~3930

\re
\ 8. Poirier J. \ et al. \ 1999, Proceedings of the 26th ICRC 4,172

\re
\ 9. Poirier J. \ et al., \ 1999, Proceedings of the 26th ICRC 5, 304

\re
\ 10. Poirier J., \ et al.,\ 2001, Proceedings of the 27th ICRC, 602

\re
\ 11. Poirier J., \ et al.,\ 2002, Physical Review D 67, 042001

\re
\ 12. Poirier J., \ et al., \ 2003, Proceedings of the 28th ICRC

\re
\ 13. Poirier J., Gress J, Lin T \ 1999, Proceedings of the 26th ICRC 2, 64

\re
\ 14. Poirier J., Roesler S., Fass\`o A., \ 2002, Astroparticle Physics 17, 441

\endofpaper
\end{document}